\documentclass{aa}
\usepackage{epsf}

\begin{document}


\newcommand{\pF}{\mbox{$p_{\mbox{\raisebox{-0.3ex}{\scriptsize F}}}$}}  
\newcommand{\vph}[1]{\mbox{$\vphantom{#1}$}}  
\newcommand{\kB}{\mbox{$k_{\rm B}$}}           
\newcommand{\vF}{\mbox{$v_{\mbox{\raisebox{-0.3ex}{\scriptsize F}}}$}}  
\renewcommand{\arraystretch}{1.1}
\newcommand{\dd}{{\rm d}}


\title{What We Can Learn from Observations of Cooling Neutron Stars}
\author{
       D.~G.~Yakovlev \inst{1} 
\and
       P.~Haensel \inst{2} 
}
\institute{
        Ioffe Physical Technical Institute, Politekhnicheskaya 26,
        194021 St.-Petersburg, Russia, {\it yak@astro.ioffe.ru} \\
\and
        Copernicus Astronomical Center,
        Bartycka 18, 00-716 Warsaw, Poland, {\it haensel@camk.edu.pl}
}

\date{}
\offprints{D.~G.~Yakovlev: yak@astro.ioffe.rssi.ru}

\titlerunning{Observations and theory of cooling neutron stars}
\authorrunning{D.~G.~Yakovlev \& P.\ Haensel}

\abstract{
A generic toy model of a cooling neutron star (NS) is used to analyze
cooling of NSs with nucleon and exotic
compositions of the cores. The model contains the
parameters which specify the levels of slow and
fast neutrino emission as well as the lower and
upper densities of the layer where  the slow
emission transforms into the fast one. The prospects to constrain
these parameters from the present and future observations
of isolated middle-aged NSs are discussed.
}

\maketitle


\section{Introduction}

Neutron stars (NSs) are compact stellar objects
which contain the matter of supranuclear density in their
cores. The equation of state
(EOS) of this matter cannot be calculated unambiguously
(e.g., Lattimer \& Prakash \cite{lp01}).
Instead, there are many theoretical
models which give a wide scatter of EOSs,
from soft to stiff ones, with
standard nucleon/hyperon or exotic compositions of matter.
The nucleon matter consists mainly of
neutrons (n) with an admixture of protons and
electrons (p and e) and possibly muons.
Some model EOSs predict also the appearance of hyperons. 
The exotic matter may contain
pion condensates, kaon condensates, or quarks
(or a mixture of these components). 
The indicated models are still almost not constrained by
the observations of NSs.

In this paper we discuss the ability to constrain
the EOS in the NS cores by confronting
the observations of thermal emission from isolated NSs with the
theory of NS cooling.
In the last three
decades the theory has been compared with the observations
by many authors (e.g., Page \cite{page98a})
but the problem is complicated and the main
features are still unknown
(too many factors are involved, such as 
neutrino emission mechanisms, superfluidity of
baryon components of matter; see, e.g.,
Yakovlev et al.\ \cite{ygkp02}). 
Thus, at this stage it is sufficient to
be general
rather than accurate and use a simple toy model
of cooling NSs. It will allow us to formulate
the current status of the problem (Sects.\ 4 and 5) without complicated
simulations. 

\section{A toy model of a cooling NS}

Let us formulate our toy model.

A middle-aged NS ($t \la 10^5$ yr)
cools mainly via neutrino emission
from its core (the region of densities
$\rho \ga 1.5 \times 10^{14}$ g cm$^{-3}$).
We assume that the core
can be subdivided into three zones:
the {\it outer} zone, $\rho< \rho_{\rm s}$;
the {\it transition} zone, $\rho_{\rm s} \leq \rho < \rho_{\rm f}$;
and the {\it inner} zone, $\rho \geq \rho_{\rm f}$.
If the central stellar density
$\rho_{\rm c} \leq \rho_{\rm s}$, two last zones are absent.

In the outer zone we assume a
{\it slow} neutrino
emission while in the inner zone we assume a
{\it fast} emission with the neutrino emissivity
$Q_\nu$ (erg s$^{-1}$ cm$^{-3}$):
\begin{equation}
     Q_\nu^{\rm slow}(\rho \leq \rho_{\rm s})=Q_{\rm s} T_9^8,\qquad
     Q_\nu^{\rm fast}(\rho \geq \rho_{\rm f})=Q_{\rm f} T_9^6.
\label{Qs}
\end{equation}
Here, $T_9$ is the internal stellar temperature $T$
expressed in $10^9$ K, while $Q_{\rm s}$ and $Q_{\rm f}$
are constants.
For simplicity, we use the linear interpolation in $\rho$ between
$Q_\nu^{\rm slow}$ and $Q_\nu^{\rm fast}$ in the transition zone.
This approximation, although oversimplified, seems to be
sufficient for exploring the main features
of the transition zone -- its position and thickness
(which are presently unknown).

The proposed {\it generic} description of $Q_\nu$
covers a number of {\it physical} model EOSs
of nucleon and exotic supranuclear
matter with different leading neutrino processes
collected in Tables 1 and 2. In these tables,
N is a nucleon (n or p);
$\nu$ and $\bar{\nu}$ are neutrino and antineutrino;
q is a quasinucleon (mixed n and p states); u and d are quarks.

For instance, $Q_{\rm s}$ can describe
modified Urca (Murca) process
in nonsuperfluid nucleon matter,
or weaker
NN-bremsstrahlung (e.g., 
nn-bremsstrahlung if Murca is
suppressed by a strong proton superfluidity
as considered by Kaminker et al.\ \cite{khy01,kyg02}).
The factor $Q_{\rm f}$ can describe the processes
of fast neutrino emission:
a powerful direct Urca (Durca) process in nucleon matter
(Lattimer et al.\ \cite{lpph91}) or
weaker (but nevertheless strong) Durca-like
processes in exotic phases of
matter (pion condensed, kaon condensed,
or quark matter) as reviewed, e.g., by Pethick (\cite{pethick92}).
The neutrino emission from hyperon matter is qualitatively
the same as from nucleon matter.
The bottom line of Table 2
refers to nonsuperfluid quark matter in NS cores.

\newcommand{\rrr}{\rule{0cm}{0.2cm}}

\begin{table}[t]
\caption{Main processes of slow neutrino emission
in nucleon matter: Murca and
bremsstrahlung (brems)}
\begin{center}
  \begin{tabular}{|lll|}
  \hline
  Process   &    &  $Q_{\rm s}$, erg cm$^{-3 \rrr}$ s$^{-1}$ \\
  \hline
  Murca &
  ${\rm nN \to pN e \bar{\nu} \quad
   pN e \to nN \nu } $ &
  $\quad 10^{20 \rrr}-3 \times 10^{21}$  \\
  Brems. &
  ${\rm NN \to NN  \nu \bar{\nu}}$  &
  $\quad 10^{19 \rrr}-10^{20}$\\
   \hline
\end{tabular}
\label{tab-nucore-slow}
\end{center}
\end{table}

The transition zone mimics a onset of the fast neutrino
emission with growing $\rho$. In nonsuperfluid matter, the lower density
$\rho_{\rm s}$ is a threshold density of the fast
emission; the threshold is usually sharp, i.e.,
$\rho_{\rm f} \approx \rho_{\rm s}$.
In realistic models of superfluid matter, the fast emission 
turns on gradually, and the transition zone may be broader 
(e.g., Yakovlev et al.\  \cite{ygkp02}).
The broadening is caused by superfluidity provided its
strength is high at the formal (nonsuperfluid) fast emission threshold
and decreases at higher $\rho$. Then the superfluidity 
strongly suppresses the fast emission at the formal threshold
and starts to open it at some higher
$\rho=\rho_{\rm s}$, where the superfluid suppression ceases
to be very strong. It fully opens the fast emission
at still higher $\rho=\rho_{\rm f}$, where the superfluid
suppression is almost completely removed. 

To be specific, we adopt a density profile within the
NS in the form: $\rho(r)=\rho_{\rm c}[1-(r/R)^2]$,
where $R$ is the NS radius.
According to Lattimer \& Prakash (\cite{lp01}), this
is a reasonable approximation for many realistic
NS models. Then the NS 
(gravitational) mass is $M=8 \pi \rho_{\rm c}R^3/15$.

To follow the NS cooling
we solve the equation of thermal
balance in the approximation of isothermal interior
(e.g., Glen \& Sutherland \cite{gs80}):
\begin{equation}
   C(T_i) \, {{\rm d}T_i \over {\rm d}t}=
       -L_\nu^\infty(T_i)-L_\gamma^\infty(T_{\rm s}),
\label{therm-balance}
\end{equation}
where $T_{\rm s}$ is the effective surface temperature,
$T_i(t)=T(r,t) \, {\rm e}^\Phi$ is the redshifted internal
temperature which is constant throughout the isothermal
interior ($\rho \ga \rho_{\rm b} \sim 10^{10}$ g cm$^{-3}$) 
with account for the
effects of General Relativity; $T(r,t)$ is the local
internal temperature of matter, and $\Phi(r)$ is the metric
function (describing gravitational redshift). 
Furthermore, $C$ is the total NS heat capacity,
$L_\gamma^\infty=4 \pi \sigma T_{\rm s}^4 \, R^2 \,
(1-r_{\rm g}/R)$ is the photon surface luminosity
as detected by a distant observer ($r_{\rm g}=2GM/c^2$
being the gravitational radius), and $L_\gamma^\infty$
is the redshifted neutrino luminosity.
We have 
\begin{equation}
  C(T_i)= \int {\rm d}V\; c(T), \qquad
    L^\infty_\nu(T_i) = \int {\rm d}V \;  
      Q_\nu(T) \,{\rm e}^{2 \Phi}, 
\label{L}
\end{equation}
where $c(T)$ is the heat capacity per unit volume,
${\rm d}V$ is an element of proper volume;
integration is done over the NS interior.
In addition, we
introduce $T_{\rm s}^\infty=T_{\rm s} \; \sqrt{1-r_{\rm g}/R}$,
the effective surface temperature detected by a distant
observer.

Taking the simplicity of the toy model,
we assume ${\rm d}V=4 \pi r^2\,{\rm d}r$ (flat space)
and ${\rm e}^\Phi=\sqrt{1-r_{\rm g}/R}$ (constant
redshift) in the NS interior.
With the above assumptions on $Q_\nu(T)$, the function
$L_\nu^\infty(T)$ is calculated in an analytic form.

We employ $c(T)=c_0(T)\,f_{\rm C}$, where $c_0(T)$ is the heat capacity
of degenerate baryon matter composed of
one particle species of number density $n=\rho/m_{\rm N}$
($m_{\rm N}$ being the bare nucleon mass) and
effective mass $m^*_{\rm N}=0.7 \, m_{\rm N}$;
$f_{\rm C}$ is the parameter introduced 
to absorb the drawbacks of our toy model and to
account for the effects of other particles and superfluidity.
The total NS heat capacity is then also evaluated in an analytic form:
$C=4 \pi \times 0.2293\, R^3 \,c_{\rm c}(T)$, where
$c_{\rm c}(T)$ is the specific heat 
at the stellar center. 
In the NS models with $Q_{\rm s}>10^{20}$
(appropriate for nonsuperfluid nucleons, see above)
we set $f_{\rm C}=1.25$ adding thus 25\% contribution
of the heat capacity of protons (Page \cite{page94}). 
In the models with $Q_{\rm s}<10^{20}$ 
(appropriate to strongly superfluid protons
at $\rho_{\rm c}<\rho_{\rm s}$)
we set $f_{\rm C}=1.0$ at $\rho_{\rm c}<\rho_{\rm s}$
and $f_{\rm C}=1.25$ at $\rho_{\rm c}>\rho_{\rm s}$. 
In fact, the value of  $f_{\rm C}$ weakly affects
the cooling curves at the neutrino cooling
stage and such variations of $f_{\rm C}$ might be
neglected: they do not change our principal results. 

We adopt
the formula of 
Potekhin et al.\ (\cite{pcy97})
to relate the internal and the surface temperatures
(for the standard NS heat blanketing envelopes made of iron).
The surface gravity in this expression has been calculated as
$g=GM \,{\rm e}^{-\Phi}/R^2$, i.e,
including
the General Relativity effects.

Thus we obtain a toy model of cooling NSs in a closed form.
The model contains {\it five parameters}: $Q_{\rm s}$, $Q_{\rm f}$,
$\rho_{\rm s}$, $\rho_{\rm f}$, and $f_{\rm C}$.
Similar models have been used, e.g.,
by Lattimer et al.\ (\cite{lvpp94}) and Page (\cite{page98a,page98b})
who have studied the sensitivity of the
NS cooling to the efficiency of fast neutrino emission
(i.e., to variations of $Q_{\rm f}$).
They implemented the generic model of $Q_\nu^{\rm fast}(T)$
into exact cooling codes.
Our toy model is much simpler but it enables us to make
a general view of the problem (by varying additionally
$Q_{\rm s}$, $\rho_{\rm s}$, and $\rho_{\rm f}$)
and mimic thus a number of physical phenomena
without complicated computation.

\section{Cooling models}

For an adopted EOS of dense matter
(for a set of five parameters, in our case) we can construct
a sequence of NS models with different $\rho_{\rm c}$
(i.e., different $M$). We can 
mimic the stiffness of
EOSs by choosing different mass-radius relations.
For simplicity, we take $R=12$ km for all models
(the approximation of constant $R$ may hold in a wide
range of $M$ for a number of EOSs, see Lattimer \& Prakash \cite{lp01}).
The main results will be the same for more
realistic mass-radius relations. We will vary $\rho_{\rm c}$
from $7 \times 10^{14}$ g cm$^{-3}$ to $1.4 \times 10^{15}$
g cm$^{-3}$ varying thus $M$
from 1.02 M$_\odot$ to 2.04 M$_\odot$.

\begin{table}[t]
\caption{Leading processes of fast
neutrino emission
in nucleon matter and three models of exotic
         matter}
\begin{center}
  \begin{tabular}{|lll|}
  \hline
  Model              & Process             &
        $Q_{\rm f}$, erg cm$^{-3 \rrr}$ s$^{-1}$ \\
  \hline
  Nucleon matter &
  ${\rm n \to p e \bar{\nu} \quad
   p e \to n \nu }$ & $\quad
  10^{26 \rrr}-10^{27}$  \\
  Pion condensate &
  ${\rm q \to q e \bar{\nu} \quad
   q e \to q {\nu} } $ & $ \quad
  10^{23 \rrr}-10^{26}$  \\
   Kaon condensate &
   ${\rm q \to q e \bar{\nu} \quad
    q e \to q \nu } $ & $ \quad
   10^{23 \rrr}-10^{24}$ \\
   Quark matter &
   ${\rm d \to u e \bar{\nu} \quad  u e \to d \nu } $ & $  \quad
   10^{23 \rrr}-10^{24}$ \\
   \hline
\end{tabular}
\label{tab-nucore-fast}
\end{center}
\end{table}

The calculations predict three
types of cooling NSs:
({\it 1}) low-mass NSs, $M < M_{\rm s}$
($M_{\rm s}$ corresponds to the central
density $\rho_{\rm c}= \rho_{\rm s}$);
({\it 2}) medium-mass NSs,
$M_{\rm s} < M \la M_{\rm f}$
($M_{\rm f}$ corresponds to $\rho_{\rm c} \sim \rho_{\rm f}$);
and ({\it 3}) high-mass NSs, $M \ga M_{\rm f}$.
The same conclusions have been made by
Kaminker et al.\ (\cite{kyg02}) with regard to
the models of NSs with nucleon cores. 

Low-mass NSs are {\it slowly cooling} objects since their neutrino
emission is slow:
$Q_\nu=Q_\nu^{\rm slow}$. Their cooling curves are almost
insensitive to $M$ due to the
obvious reason: both, $L_\nu$ and $C$, increase with
increasing $M$ but this increase is compensated
in the ratio $L_\nu/C$ which determines the cooling
rate at the neutrino-cooling stage
($L_\nu \gg L_\gamma$).

High-mass NSs cool mainly via fast neutrino
emission, $Q_\nu=Q_\nu^{\rm fast}$, from the inner zones.
Their cooling curves
are also not too sensitive to
$M$. These {\it rapidly cooling} middle-aged stars are noticeably
colder than the slowly cooling ones.

Medium-mass NSs show the cooling
intermediate between the slow and fast ones.
While increasing $M$ from $M_{\rm s}$
to $M_{\rm f}$ and higher, we get a sequence of
cooling curves
which realize the transition from the slow to the fast
cooling regimes.

\begin{figure}
\centering
\epsfysize=80mm
\epsffile[70 215 550 680]{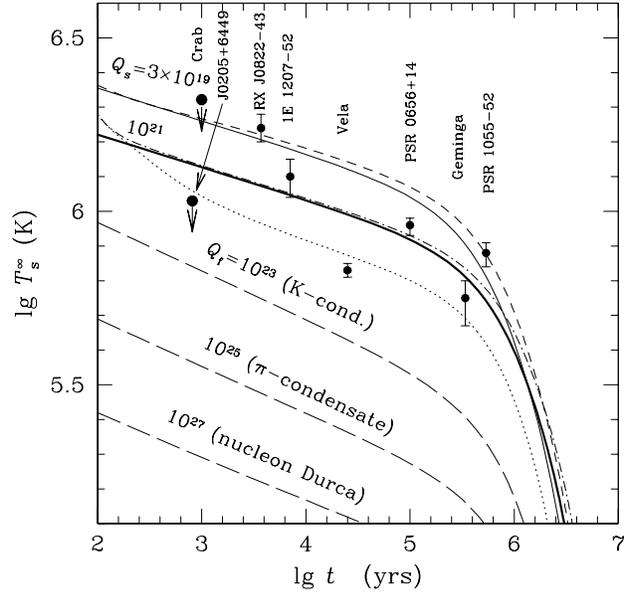}
\caption{
Observational limits on surface temperatures of eight
NSs versus toy-model cooling curves
of low-mass NSs (solid lines) with two 
slow neutrino emission levels
and high-mass NSs (long dashes) with
three fast neutrino emission levels.
Dashed-and-dot line: slow cooling of a nonsuperfluid
1.4 M$_\odot$ NS calculated with the exact cooling code;
dotted line: the same but assuming
triplet-state neutron pairing in the core,
Takatsuka (\cite{takatsuka72}) model;
short-dashes: the same but with strong proton pairing
and nonsuperfluid neutrons.
}
\label{fig1}
\end{figure}

The thick and thin solid lines in Fig.\ \ref{fig1} display typical cooling
curves of low-mass NSs
($\rho_{\rm c}= 8 \times 10^{14}$ g cm$^{-3}$, $M=1.16 \, {\rm M}_\odot$)
with two levels
of the slow neutrino emission, $Q_{\rm s}$.
The long-dashed lines show cooling curves
for high-mass NSs ($\rho_{\rm c}=  1.4 \times 10^{15}$
g cm$^{-3}$, $M=2.04 \, {\rm M}_\odot$,
$\rho_{\rm s}=8 \times 10^{14}$ g cm$^{-3}$,
$\rho_{\rm f}=10^{15}$ g cm$^{-3}$)
with three levels of the fast neutrino
emission, $Q_{\rm f}=10^{23}$, $10^{25}$ and $10^{27}$,
which roughly correspond to
NSs with kaon condensates, pion condensates, and nucleon matter
with Durca process in NS cores. NSs with hyperon cores
are expected to cool at about the same rate as NSs with
nucleon cores.
One can see great difference of cooling scenarios at various
$Q_{\rm s}$ and $Q_{\rm f}$.

The thick solid curve
($Q_{\rm s}= 10^{21}$) is very close to
those obtained (e.g.,  Page \cite{page98a,page98b},
Kaminker et al.\ \cite{kyg02})
for low-mass non-superfluid NSs with the nucleon cores
using exact cooling codes.
It may be regarded as the {\it basic slow-cooling curve}.

For comparison, we present three cooling curves
calculated with an exact cooling code for a 1.4 M$_\odot$
NS (the nucleon core with forbidden
Durca process, $\rho_{\rm c}=1.22 \times 10^{15}$ g cm$^{-3}$,
$R$=11.65 km, EOS B in notations of Kaminker et al.\
\cite{kyg02}). The dot-and-dashed curve refers to
a nonsuperfluid star and agrees with the basic slow-cooling
curve. The short-dashed curve is for a NS with strong
proton superfluidity (which switches off the proton heat capacity
and the neutrino reactions involving protons);
it agrees with the upper solid curve. By varying 
the heat capacity parameter $f_{\rm C}$
of our toy model, we could get even better agreement
with the two exact cooling curves indicated above. 
At $t \la 100$ yr the exact code accurately describes
the NS thermal relaxation which is not
reproduced by the toy model.
At these $t$, the exact cooling curves
noticeably deviate from the toy-model ones. The third exact
curve is explained later.

\begin{figure}
\centering
\epsfysize=80mm
\epsffile[70 215 550 680]{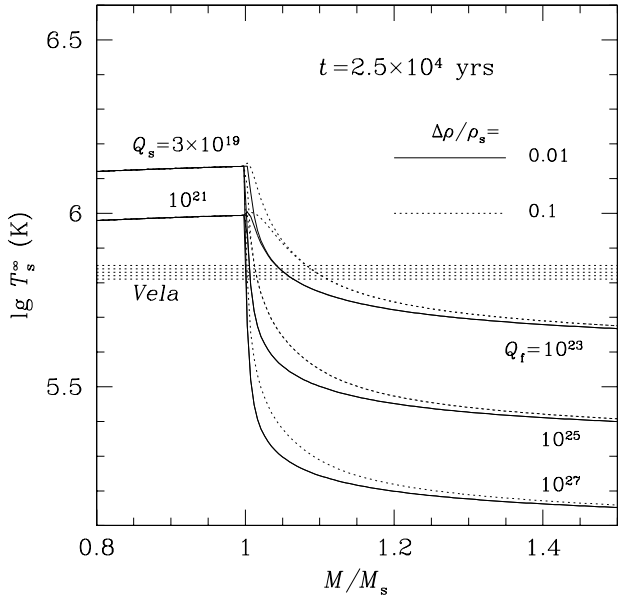}
\caption{
Transition from slow to fast cooling
with increasing NS mass over
the threshold value $M_{\rm s}$
for the same values of $Q_{\rm s}$ and $Q_{\rm f}$
as in Fig.\ \ref{fig1}
and two transition density intervals,
$\Delta \rho/\rho_{\rm s}=0.01$ or 0.1,
at a fixed NS age $t=25000$ yrs.
Shaded strip: observational limits for the Vela pulsar.
}
\label{fig2}
\end{figure}

Figure \ref{fig2} shows the transition from  slow to  fast cooling
with increasing $M$ for the same $Q_{\rm s}$ and $Q_{\rm f}$,
$\rho_{\rm s}=8 \times 10^{14}$ g cm$^{-3}$
($M_{\rm s}=1.16 \, {\rm M}_\odot$),
and two relative density widths of the transition
zone
$\Delta \rho/\rho_{\rm s} \equiv
(\rho_{\rm f}-\rho_{\rm s})/\rho_{\rm s}=0.01$
or 0.1. The first value of $\Delta \rho/\rho_{\rm s}$ 
corresponds to a sharp
threshold of the fast emission, while the second value is
appropriate to a smooth threshold.
The NS age is fixed, $t=25000$ yrs
(the age of the Vela pulsar,
Lyne et al.\ \cite{lyneetal96}).
The main features
of Fig.\ \ref{fig2} would not change if we
varied $\rho_{\rm s}$ from $8 \times 10^{14}$
to $1.2 \times 10^{15}$ g cm$^{-3}$ ($M_{\rm s}$ from
1.16 M$_\odot$ to 1.75 M$_\odot$).
If the contrast $Q_{\rm f}/Q_{\rm s}$ between the fast and slow
emissivities is not too high 
($Q_{\rm f}/Q_{\rm s} \la 10^3$),
the transition is rather smooth
even for a very narrow density width $\Delta \rho$.
For higher contrasts,
the transition is smooth only if the density width
is not too small,
$\Delta \rho/\rho_{\rm s} \ga 0.1$.

\section{Confronting theory with observations}

Although the toy model is oversimplified,
it reproduces the main features of
exact calculations and can be confronted with
the observations of thermal emission from
cooling NSs.

The observational basis is shown in Fig.\ \ref{fig1}.
It displays
the observational values of $T_{\rm s}^\infty$
for six middle-aged isolated NSs, the same as in
Yakovlev et al.\ (\cite{ygkp02}) 
excluding RX J1856--3754 and RX J0002+62.
The latter two sources are most interesting
but the interpretation of the observed spectra and the
extraction of the surface thermal radiation is complicated
(as discussed, e.g., 
in Pavlov et al.\ \cite{pzs02}, Walter \& Lattimer \cite{wl02},
and in references therein with regard to RX J1856--3754;
as commented by Pavlov \cite{pavlov02} with regard to RX J0002+62). 
The objects presented in Fig.\ \ref{fig1} are:
RX J0822--43, 
1E 1207--52, 
Vela, 
PSR 0656+14, 
Geminga, 
and PSR 1055--52. 
The data on
$T_{\rm s}^\infty$
and $t$ for these NSs are taken from the sources cited in
Kaminker et al.\ (\cite{kyg02}).
We display also
the upper limit of $T_{\rm s}^\infty$ for the Crab pulsar
(Tennant et al.\  \cite{crab}) and
for PSR J0205+6449 in the supernova remnant 3C 58
(Slane et al.\ \cite{shm02}).

The comparison of the data with the
cooling curves leads to the following
{\it generic} conclusions.

(1) The hottest observed NS, RX J0822--43,
lies {\it well above the basic slow-cooling curve} but
is compatible with the models of low-mass NSs
with $Q_{\rm s} \sim 3 \times 10^{19}$, i.e., with the {\it reduced
slow neutrino emission}.  These conclusions
have been made, e.g., by Kaminker et al.\
(\cite{khy01,kyg02}). Alternatively, one can assume
either additional reheating mechanisms in the stellar interiors
(which will complicate the theory) or
the presence of accreted NS envelopes
(which increase the electron thermal conductivity and
make the surface layers
hotter, at the neutrino cooling stage, see, e.g., 
Potekhin et al.\ \cite{pcy97} and Page \cite{page98a}).

(2) The coldest observed NSs, particularly, the
Vela and Geminga pulsars, lie {\it well
below the basic slow-cooling curve}.
They {\it require enhanced neutrino emission}
in their cores, with $Q_{\rm f} \ga 10^{23}$,
but {\it we cannot
pinpoint the nature of this enhancement}. This conclusion
has been made by several authors
(e.g., Page \cite{page98a,page98b}).
Even a weak enhancement produced
by kaon condensates would be consistent with
the data, but any stronger enhancement
produced by pion condensates or nucleon Durca
is also allowed. Thus,
future search for colder NSs would be
crucial but it may take time because
of difficulties in detecting faint sources.

(3) A rather uniform scatter of observational
points between the hottest and coldest NSs
indicates {\it a sufficiently smooth transition} from
slow to fast cooling, i.e., {\it the existence
of representative class of medium-mass} NSs.
This circumstance has been mentioned, e.g.,
by Kaminker et al. (\cite{khy01,kyg02})
with regard to the cooling of NSs with
nucleon cores.   

(4) The threshold density of fast neutrino
emission, $\rho_{\rm s}$, can be placed anywhere
in the interval from $\sim 8 \times 10^{14}$ g cm$^{-3}$
to $\sim 1.2 \times 10^{15}$ g cm$^{-3}$. For any
$\rho_{\rm s}$ from this interval we can
to build a sequence of models of medium-mass NSs. Tuning their
masses, we could explain the data.
Adopting various $\rho_{\rm s}$ and $\rho_{\rm f}$
we will attribute {\it different masses} to the same sources,
as seen from Fig.\ 2 with the Vela pulsar as an example.
Similar conclusions were made
by Kaminker et al. (\cite{khy01,kyg02})
for cooling NSs with
nucleon cores. Their transition layer
was produced by the weakening of the proton pairing
with increasing $\rho$ and the associated broadening 
of the Durca threshold. 

(5) For a not too high contrast of slow and fast neutrino
emissivities ($Q_{\rm f}/Q_{\rm s} \la 10^3$)
we can build a representative class of medium-mass NSs even 
with a narrow
density width of the transition zone, $\Delta \rho/\rho_{\rm s} \ll 1$.
For sharper contrasts, the density width has to
be sufficiently wide, $\Delta \rho / \rho_{\rm s} \ga 0.1$,
as follows from the cooling simulations
of superfluid NSs with nucleon cores 
(Kaminker et al.\ \cite{khy01,kyg02}).
One needs a wide transition zone 
to obtain a representative class of medium-mass NSs
with superfluid nucleon or pion-condensed cores.
This is possible in the presence of
a strong superfluidity in the vicinity
of the fast-emission threshold.

\section{Discussion and conclusions}

Thus, our {\it generic cooling analysis} 
(5-parameter physics input leading to the families of
cooling curves of NSs with different $M$) 
is too flexible to fix 
the nature of the fast neutrino emission and its density threshold.
Consequently, the data {\it can be 
explained by a number of physical EOSs} of dense matter.
Notice that the cooling curves are more sensitive to 
the composition of matter than
to the stiffness of the EOS (although the composition and stiffness
are actually interrelated).

For instance, Kaminker et al.\ (\cite{khy01,kyg02}) 
exploit the idea of nucleon matter in
the NS core with the onset of the Durca process at high densities. 
They assume the presence of
a strong proton superfluidity at not too high densities 
to suppress the Murca process and
broaden the Durca threshold. The suppression 
of Murca allows them to reduce the slow
neutrino emission level $Q_{\rm s}$ 
(from about $10^{21}$ to about $3 \times 10^{19}$)
and explain thus the hottest observed sources 
(Fig.\ \ref{fig1}). The broadening of
the Durca threshold ensures a representative class of 
medium-mass NSs to interpret cooler objects.

In the phenomenological approach of 
Kaminker et al.\ (\cite{khy01,kyg02}) the crucial 
$^1{\rm S}_0$ proton superfluidity is modeled by a specific density
dependence of the proton critical 
temperature $T_{\rm cp}$. Consider, for instance, a
particular realization of this model 
applied by Yakovlev et al.\ (\cite{ykhg02}) to analyze
the thermal state of 
PSR J0205+6449. 
For this
specific model (model 1p with EOS A, in notations of the authors), 
the maximum of $T_{\rm cp}$ ($\sim 7 \times 10^9$~K)  is
reached at $5.75 \times 10^{14}$ g cm$^{-3}$. 
The critical temperature $T_{\rm cp}$ decreases at higher
densities, but it is still $\approx 4.1 \times 10^9$~K 
at the Durca threshold, $\rho_{\rm D}=7.85 \times 10^{14}$
g cm$^{-3}$, and then drops to zero at
about $1.2 \, \rho_{\rm D}$.
Can $T_{\rm cp}$ be as high as $\sim 4\times 10^9$~K  
at $\rho=\rho_{\rm D}$ where
the proton fraction is about 11\%? 
In a recent paper Tsuruta et al. (\cite{tsurutaetal02})
argue that this would contradict to 
the existing microscopic models of nucleon
superfluidity, because  
the proton number density $n_{\rm p}$ 
is too large at $\rho=\rho_{\rm D}$  
to allow for the $^1{\rm S}_0$ proton pairing.
Taking the model of Yakovlev et al.\ (\cite{ykhg02})
mentioned above
we get 
$n_{\rm p}=0.047$ fm$^{-3}\approx 0.3 \,n_0 $ at
$\rho=\rho_{\rm D}$, i.e., noticeably smaller than 
$n_0=0.16$ fm$^{-3}$, the nucleon number density
in saturated nuclear matter. 
For the same number density of neutrons in neutron matter, the
critical temperature of the $^1{\rm S}_0$ neutron superfluidity is
$\sim 1.3\times 10^{10}~$K if a realistic bare NN interaction
is used (Lombardo \& Schulze
\cite{ls01}). 
The medium effects (polarization effects, self-energy corrections) can
decrease it to $7 \times 10^9~$K
or lower, depending on a particular model. 
In the case of the proton component of the
neutron-star matter the np interaction can decrease the proton effective
mass, lowering further $T_{\rm cp}$. However,
this does not mean that the value $T_{\rm cp}(\rho_{\rm D})
\ga 7 \times 10^{9}~$ K, followed by a rapid drop of $T_{\rm cp}$ 
with increasing $\rho$, is ruled out by contemporary
microscopic theories.
In contrast to the case of neutron matter at $n_{\rm n} < n_0$,
reliable calculation of $T_{\rm cp}(\rho_{\rm D})$ in
the neutron-star matter, starting  from
a realistic NN interaction and including the
medium effects, remains still
a challenge for the many-body theory.
Therefore, we think that the use of a ``minimal model'' (npe matter
with Durca acting at $\rho>\rho_{\rm D}$, 
combined with the appropriate nucleon
superfluidity) remains a valid first-step 
approach.

Notice that our general assumption on the neutrino
emissivity $Q_\nu(T,\rho)$ would be violated
in the presence of ($^3$P$_2$) 
neutron superfluidity in the NS core
with a density dependent
critical temperature $T_{\rm cn}(\rho)$
which has the maximum in the range from
$\sim 10^8$ to $\sim 2 \times 10^9$ K (see, e.g.,
Kaminker et al.\ \cite{kyg02}).
The neutrino emission due to Cooper pairing
of neutrons will then be so strong that it will
initiate a really fast cooling even at
$M < M_{\rm s}$,
violating the interpretation of relatively
hot and old sources, first of all PSR 1055--52.
As an example, by the dotted line 
in Fig.\ \ref{fig1} we show
the cooling of 1.4 M$_\odot$ NS
with the nucleon core and forbidden Durca process
in the presence of a neutron superfluidity
(model of Takatsuka \cite{takatsuka72}, with
maximum $T_{\rm c}$ of about $9 \times 10^8$ K at
$\rho \approx 5 \times 10^{14}$ g cm$^{-3}$).
One can see that NSs with this superfluidity would be too cold
to explain the data.

Another physical model of cooling NSs was
presented by Tsuruta
et al.\ (\cite{tsurutaetal02}). It is
based on pion condensation at supranuclear
densities exploiting similar idea: slow
cooling of low-mass NSs and faster cooling
of massive NSs. Their cooling curve of low-mass NS (1.2 M$_\odot$,
nucleon core, forbidden Durca process)
agrees with our basic slow-cooling curve (after
translating our curve, $T_{\rm s}^\infty(t)$, to their
format, $L_\gamma^\infty(t)$). 
Note that Tsuruta et al.\ (\cite{tsurutaetal02})
misplaced the positions of some observational
data in their figure. Most important is RX J0822--43. They
(as well as we) take
the data from Zavlin et al.\ (\cite{ztp99}) who
give $L_\gamma^\infty=(7.1-10.1)\times 10^{33}$ erg s$^{-1}$
(${\rm lg}\, L_\gamma^\infty=33.85-34.00$) while Tsuruta et al.\
present ${\rm lg}\,L_\gamma^\infty \approx 33.57-33.97$. Thus,
RX J0822--43 is sufficiently warm and cannot be explained
with the basic slow-cooling model (Fig.\ \ref{fig1}).  
Moreover, according to Tsuruta et al., they
employ the model neutron superfluidity of Takatsuka (\cite{takatsuka72}).
Therefore, their curve (if properly calculated) should resemble our
dotted curve, and their model would then disagree
with the number of observational limits. 
To avoid this disagreement one can
change the model of nucleon superfluidity.
A natural model of rather strong proton superfluidity
and weak neutron superfluidity at $\rho \sim (3-8)\times 10^{14}$
g cm$^{-3}$ considered by Kaminker et al.\ (\cite{khy01,kyg02})
seems to be most suitable. 

\begin{acknowledgements}
The authors are grateful to anonymous referee,
A.D.\ Kaminker, and K.P. Levenfish
for useful critical remarks.
This work was supported in part by the
RFBR (grants 02-02-17668 and 03-07-90200)
and KBN (grant 5 P03D 020 20).
\end{acknowledgements}


\begin{thebibliography}{22}

\bibitem[1980]{gs80} 
Glen,~G., \& Sutherland,~P. 1980,
ApJ 239, 671


\bibitem[2001]{khy01}
Kaminker, A.~D., Haensel,~P., \& Yakovlev, D.~G. 2001,
A\&A 373, L17 


\bibitem[2002]{kyg02}
Kaminker, A.~D., Yakovlev, D.~G., \& Gnedin, O.~Y. 2002,
A\&A 383, 1076


\bibitem[2001]{lp01}
Lattimer, J.~M., \& Prakash, M. 2001,
ApJ 550, 426


\bibitem[1991]{lpph91}
Lattimer, J.~M., Pethick, C.~J.,  Prakash, M., \& Haensel, P. 1991,
Phys.\ Rev.\ Lett.\ 66, 2701

\bibitem[1994]{lvpp94}
Lattimer, J.~M.,  Van Riper, K.~A., Prakash, M., \& Prakash, M. 1994,
ApJ 425, 802

\bibitem[2001]{ls01}
Lombardo, U., Schulze, H.-J., 2001,
in: Physics of Neutron Star Interiors,
eds D.\ Blaschke, N.~K.\ Glendenning, \& A.\ Sedrakian
(Lecture Notes in Physics, Springer, Berlin) p.~30 

\bibitem[1996]{lyneetal96}
Lyne, A.~G., Pritchard, R.~S., Graham-Smith, F., \& Camilo, F. 1996,
Nature 381, 497

\bibitem[1994]{page94}
Page,~D. 1994, ApJ 428, 250

\bibitem[1998a]{page98a}
Page,~D. 1998a,
in: The Many Faces of Neutron Stars,
eds R.\ Buccheri, J.\ van Paradijs, \& M.~A.\ Alpar
(Kluwer, Dordrecht) p.\ 539

\bibitem[1998b]{page98b}
Page,~D. 1998b,
in: Neutron Stars and Pulsars,
eds N.\ Shibazaki,
N.\ Kawai, S.\ Shibata, \& T.\ Kifune
(Universal Academy Press, Tokyo) p.~183

\bibitem[2002]{pavlov02}
Pavlov,~G.~G. 2002, private communication

\bibitem[2002]{pzs02}
Pavlov,~G.~G., Zavlin,~V.~E., \& Sanwal,~D. 2002,
in: Proc.\ of 270 Heraeus Seminar on
Neutron Stars, Pulsars and Supernova Remnants,
eds W.\ Becker, H.\ Lesh, \& J.\ Tr\"umper (MPE,
Garching) p.~273

\bibitem[1992]{pethick92}
Pethick, C.~J. 1992, Rev.\ Mod.\ Phys. 64, 1133

\bibitem[1997]{pcy97}
Potekhin, A.~Y., Chabrier, G., \& Yakovlev, D.~G.
1997, A\&A 323, 415

\bibitem[2002]{shm02}
Slane, P., Helfand, D.~J., \& Murray, S.~S. 2002,
ApJ (Letters) 571, L45

\bibitem[1972]{takatsuka72}
Takatsuka, T. 1972,
Prog.\ Theor.\ Phys.\ 48, 1517

\bibitem[2001]{crab}
Tennant, A.~F., Becker, W., Juda, M., Elsner, R.~F.,
Kolodziejczak, J.~J., Murray, S.~S., O'Dell, S.~L.,
Paerels, F., Swartz, D.~A., Shibazaki, N., \& Weisskopf, M.~C.
2001, ApJ (Letters) 544, L173

\bibitem[2002]{tsurutaetal02}
Tsuruta, S., Teter, M.~A., Takatsuka, T., Tatsumi, T., \&
Tamagaki, R. 2002,
ApJ (Letters) 571, L143

\bibitem[2002]{wl02}
Walter, F.~M., \& Lattimer, J. 2002, ApJ (Letters) 576, L145


\bibitem[2001]{ykgh01}
Yakovlev, D.~G., Kaminker, A.~D., Gnedin, O.~Y., \& Haensel, P. 2001,
Phys.\ Rep.\ 354, 1

\bibitem[2002]{ygkp02}
Yakovlev, D.~G., Gnedin, O.~Y., Kaminker, A.~D., \& Potekhin, A.~Y.
2002,
in:
Proc.\ of 270 Heraeus Seminar on
Neutron Stars, Pulsars and Supernova Remnants,
eds W.\ Becker, H.\ Lesh, \& J.\ Tr\"umper (MPE,
Garching) p.~287

\bibitem[2002]{ykhg02}
Yakovlev, D.~G., Kaminker, A.~D., Haensel, P., \& Gnedin, O.~Y.
2002, A\&A 389, L24

\bibitem[1999]{ztp99}
Zavlin, V.~E., Tr\"umper, J., \& Pavlov, G.~G. 1999, ApJ 525, 959

\end{thebibliography}
\end{document}